\pdfoutput=1

\documentclass[11pt,twoside,a4paper,cmspaper,final,collab]{cms-tdr}
\def\svnVersion{198241}\def\cmsCernNo{2012-361}\def\cmsCernDate{\today}\def\cmsMessage{Published in JHEP as \href{http://dx.doi.org/10.1007/JHEP03(2013)037}{\doi{10.1007/JHEP03(2013)037}}. Includes erratum \href{http://dx.doi.org/10.1007/JHEP07(2013)041}{\doi{10.1007/JHEP07(2013)041}}.}

\begin{document}\cmsNoteHeader{SUS-12-017}

\hyphenation{had-ron-i-za-tion}
\hyphenation{cal-or-i-me-ter}
\hyphenation{de-vices}

\RCS$Revision: 196656 $
\RCS$HeadURL: svn+ssh://alverson@svn.cern.ch/reps/tdr2/papers/SUS-12-017/trunk/SUS-12-017.tex $
\RCS$Id: SUS-12-017.tex 196656 2013-07-17 03:06:56Z claudioc $
\newlength\cmsFigWidth
\ifthenelse{\boolean{cms@external}}{\setlength\cmsFigWidth{0.85\columnwidth}}{\setlength\cmsFigWidth{0.4\textwidth}}
\ifthenelse{\boolean{cms@external}}{\providecommand{\cmsLeft}{top}}{\providecommand{\cmsLeft}{left}}
\ifthenelse{\boolean{cms@external}}{\providecommand{\cmsRight}{bottom}}{\providecommand{\cmsRight}{right}}
\cmsNoteHeader{SUS-12-017} 
\title{Search for new physics in events with same-sign dileptons
and b jets in pp collisions at $\sqrt{s}=8$\TeV}

\date{\today}

\abstract{
A search for new physics is performed using events with isolated
same-sign leptons and at least two bottom-quark jets
in the final state.
Results are based on a
sample of proton-proton
collisions collected at a center-of-mass energy of 8\TeV with the
CMS detector and corresponding to an integrated luminosity of 10.5\fbinv.
No excess above the standard model background is observed.
Upper limits are set on the number of events
from non-standard-model sources and
are used to constrain a number of new physics models.
Information on acceptance and efficiencies is also provided so that the
results can be used to confront an even broader class of new physics models.}

\hypersetup{%
pdfauthor={CMS Collaboration},%
pdftitle={Search for new physics in events with same-sign dileptons
and b jets in pp collisions at sqrt(s) = 8 TeV},%
pdfsubject={CMS},%
pdfkeywords={CMS, physics, supersymmetry}}

\maketitle 

\section{Introduction\label{sec:introduction}}

Events with same-sign, isolated leptons (\Pe\ or \Pgm) from
standard model (SM) processes in proton-proton collisions
are extremely rare.  As a result, searches for anomalous
production of same-sign dileptons can be very sensitive to new physics
contributions.  These include
supersymmetry (SUSY)~\cite{Barnett:1993ea,Guchait:1994zk,Baer:1995va},
universal extra dimensions~\cite{Cheng:2002ab},
pair production of $T_{5/3}$ particles (fermionic partners of the top
quark)~\cite{Contino:2008hi},
heavy Majorana neutrinos~\cite{Almeida:1997em},
and same-sign top-pair production~\cite{Han:2009,fcnczprime}.
More specifically, the signature of same-sign dileptons accompanied by bottom
quarks can arise from SUSY processes where third-generation quark
superpartners are lighter than other
squarks~\cite{naturalness1,naturalness2,naturalness3,naturalness4,Csaki:2012fh},
resulting in an abundance of top and bottom quarks produced in cascade decays.
These scenarios are particularly interesting for two reasons.
First, because they
have not been excluded by early SUSY searches at the Large Hadron
Collider (LHC);
second, because relatively light bottom and top squarks are
necessary if SUSY is to be the ``natural'', (i.e., not fine-tuned),
solution to the gauge
hierarchy problem (see for example the discussion in
Ref.~\cite{naturalness4}).

In this Paper we present the result of a search for new physics in
events with same-sign isolated dileptons and two or more bottom-quark jets
(b jets).  The search was performed on a dataset
corresponding to an integrated luminosity of 10.5\fbinv of
proton-proton
(\Pp \Pp) collisions collected at a center-of-mass energy of 8\TeV
with the Compact Muon Solenoid (CMS)~\cite{JINST} detector
at the LHC.
This search is an extension of the search at
7\TeV described in Ref.~\cite{Chatrchyan:2012sa}.

In the analysis, we select events with two
isolated, high transverse momentum (\pt), same-sign leptons
(\Pe\ or \Pgm),
and at least two jets identified as b jets (b-tagged jets).
The numbers of observed events
in signal regions defined by additional
requirements on missing transverse energy (\MET), the number of
b-tagged jets, and the scalar sum of jet \pt (\HT), are compared
with SM
expectations.
Having found
no evidence for a beyond-the-SM contribution to the event counts,
we set limits on the number of events from new physics in the
event sample.  These limits are then
translated into a limit
on the same-sign top-pair production
cross section and are used to bound the parameters
of other new physics models.  Finally, we include
additional information on the event selection efficiencies
so that our results can be used to test models of new physics
not explicitly considered here.
\section{The CMS detector\label{sec:detector}}
The central feature of the CMS apparatus is a superconducting solenoid,
of 6\unit{m} internal diameter, providing a magnetic
field of 3.8\unit{T}.
The CMS experiment uses a right-handed coordinate system, with the origin defined to be the nominal interaction point,
the $x$ axis pointing to the center of the LHC ring,
the $y$ axis pointing up (perpendicular to the LHC plane),
and the $z$ axis pointing in the counterclockwise beam direction.
The polar angle $\theta$ is measured from the positive $z$ axis and the azimuthal angle $\phi$
is measured in the $x$-$y$ (transverse) plane.
The pseudorapidity $\eta$ is defined as $\eta = - \ln{[\tan{(\theta/2)}]}$.
Within the field volume are a silicon pixel and strip tracker, a crystal electromagnetic calorimeter,
and a brass-scintillator hadron calorimeter.
Muons are measured in gas-ionization detectors embedded in the steel return yoke.
Full coverage is provided by the tracker, calorimeters, and the muon detectors within $\abs{\eta}< 2.4$.
In addition to the barrel and endcap calorimeters up to $\abs{\eta}=3$,
CMS has extensive forward calorimetry reaching $\abs{\eta} \lesssim 5$.
A more detailed description of the
CMS apparatus can be found in Ref.~\cite{JINST}.
\section{Event selection and Monte Carlo simulation}
\label{sec:evsel}
The event selection is essentially the same as for the 7\TeV
analysis~\cite{Chatrchyan:2012sa}.
We require two isolated same-sign leptons (\Pe\ or \Pgm),
consistent with originating from the same vertex and
with pair invariant mass above 8\GeV, and
at least two b-tagged jets.
Events are collected with dilepton triggers.
The kinematic requirements on leptons and jets
are summarized
in Table~\ref{tab:kin}.
Events with a third lepton are rejected if the lepton forms an
opposite-sign same-flavor pair with one of the first two leptons,
 and if the invariant mass $m_{\ell \ell}$
of the pair satisfies
$m_{\ell \ell} < 12$\GeV ($\pt > 5$\GeV)
or $76 < m_{\ell \ell} < 106$\GeV ($\pt > 10$\GeV),
where the conditions in parentheses show the respective
$\pt$ requirement on the third lepton.
These requirements are designed to minimize backgrounds from
events with $\gamma^* \to \ell^+ \ell^-$,
low-mass bound-state, and multi-boson
($\PW\cPZ$ and $\cPZ\cPZ$, as well as tribosons) production.

\begin{table}[h]
\begin{center}
\topcaption{\label{tab:kin} Kinematic requirements for leptons and
jets in this analysis. }
\begin{tabular}{|l|c|c|}\hline
              & $\pt$          & $\eta$ \\ \hline
electrons     & $\pt > 20$\GeV & $|\eta| < 1.442$ or $1.566 < |\eta| < 2.4$ \\
muons         & $\pt > 20$\GeV & $|\eta| < 2.4$ \\
jets          & $\pt > 40$\GeV & $|\eta| < 2.4$ \\
b-tagged jets & $\pt > 40$\GeV & $|\eta| < 2.4$ \\ \hline
\end{tabular}
\end{center}
\end{table}

With respect to Ref.~\cite{Chatrchyan:2012sa},
there have been a few
modifications in the lepton identification and
isolation requirements. The main difference is that
lepton isolation is now computed with particle-flow
information~\cite{PFPAS}
rather than with charged tracks and calorimeter
deposits, and an event-by-event
correction is made to account for the effect
of multiple \Pp \Pp\ interactions in the same bunch crossing
(pileup); the mean number of interactions
in the same crossing is 15.
This correction consists of subtracting from the measured 
sum-$P_T$ in the isolation cone the estimated contribution from
pileup. 
This contribution is of order 1\GeV, and it is calculated based on a 
measurement of the hadronic event activity not associated with the
\Pp \Pp\ interaction that produced the leptons.  
As a result of this correction, the efficiency of the isolation
requirement for leptons from W and Z decays is almost independent of pileup.

The jet reconstruction is the same as in Ref.~\cite{Chatrchyan:2012sa},
but the b-tagging algorithm has been changed slightly.
We now use the CSV method of Ref.~\cite{CMS-PAS-BTV-11-003},
which is based on the combination of
secondary-vertex reconstruction and track-based
lifetime information.
Efficiencies for lepton identification, trigger efficiency,
and b-tagging
are summarized in Sections~\ref{sec:eff}
and~\ref{sec:outreach}.

Monte Carlo (MC) simulations including pileup are used to estimate some of the SM backgrounds
(see Section~\ref{sec:bg}), as well as to calculate the efficiency for
various new physics scenarios.
In this analysis, SM background samples and new physics signal samples
are generated with the
\MADGRAPH V5~\cite{MADGRAPH5} and
\PYTHIA V6.4~\cite{PYTHIA} programs, respectively.
Standard model processes are simulated using a
\GEANTfour-based model~\cite{Geant}
of the CMS detector; the simulation of new physics signals
is performed using the CMS fast
simulation package~\cite{Abdullin:1328345}.
All simulated events are processed with the same chain of
reconstruction programs used for collision data.
\section{Backgrounds}
\label{sec:bg}

There are three sources of SM background to potential new physics signals:
\begin{itemize}

\item ``fake leptons'', i.e., leptons from heavy-flavor decay,
misidentified hadrons, muons from light-meson decay in flight, or electrons
from unidentified photon conversions.  These are estimated from a
sample of events with at least one lepton that passes a loose
selection but fails the full set of tight identification
and isolation requirements.  The background estimation uses the
``tight-to-loose'' ratio, i.e., the probability for
a loosely identified fake lepton to also pass the full
set of requirements, as determined from studies of fake
leptons in multijet events.

\item ``charge flips'', i.e., events with opposite-sign isolated
leptons where one of the leptons is an electron and its charge is
misreconstructed due to severe bremsstrahlung in the tracker material
(this effect is negligible for muons).
Charge flips are estimated by selecting opposite-sign \Pe \Pe\
or \Pe \Pgm\
events passing the full kinematic selection, weighted by the
\pt- and $\eta$-dependent probability of electron charge
misassignment.  This probability is obtained from MC
simulation and is validated on a sample of $\cPZ \to \Pe \Pe$ events.

\item rare SM processes that yield same-sign high-$\pt$ leptons
and b jets, mostly from $\ttbar\PW$ and $\ttbar\cPZ$ production.
This background contribution is obtained from Monte Carlo simulation.

\end{itemize}

More details on the background estimation can be found in
Refs.~\cite{Chatrchyan:2012sa, sspaper2010}.
The systematic uncertainties
are 20\% for charge flips, and 50\% for both fake and rare
SM backgrounds.
The next-to-leading order $\ttbar\PW$ and
$\ttbar\cPZ$ production
cross sections
used to normalize the MC predictions are
232\unit{fb}~\cite{Campbell:2012dh}
and 208\unit{fb}~\cite{ttzNLO,Garzelli:2011is},
respectively.
\section{Event yields}
\label{sec:yields}

The search is based on comparing the event yields and background
predictions in nine
signal regions (SRs)
with different requirements on \MET, \HT, and the number of b-tagged jets.
In order to improve the
sensitivity to  $\Pp \Pp \to \cPqt \cPqt$ production,
we also define a SR with positive leptons only.
The requirements for the SRs are loosely motivated by possible new
physics scenarios.  For example,
signal regions with high \MET are most sensitive to
R-parity-conserving
SUSY models with relatively light neutralinos;
requiring
larger \HT reduces the SM background while generally
maintaining good efficiency to detect new heavy particles that are
produced via the strong interaction;
the signal region with three or more b-tagged jets is sensitive
to final states that include several b quarks.

	\begin{table}[hbt]
		\topcaption{\label{tab:SRcomb} A summary of the results of this search.
		For each signal region (SR),
		we show its distinguishing kinematic requirements, the prediction
		for the three background  (BG) components, as well as their total,
		and the observed number of events.  We also show the 95\% CL upper
                limit on the number of new physics events ($N_{UL}$) under different
               assumptions on  the signal efficiency uncertainty.
		Note that the number of jets on the first line of
		the table includes both b-tagged and non b-tagged jets.}
		\tabcolsep 2.7pt
		\begin{scriptsize}
			\begin{tabular}{|l|c|c|c|c|c|c|c|c|c|}
			\hline
                                               & SR0                    & SR1                    & SR2                    & SR3                    & SR4                    & SR5                    & SR6                    & SR7                    & SR8            \\
                        \hline
			No. of jets            & $\geq 2$               & $\geq 2$               & $\geq 2$               & $\geq 4$               & $\geq 4$               & $\geq 4$               & $\geq 4$               & $\geq 3$               & $\geq 4$       \\
			No. of btags           & $\geq 2$               & $\geq 2$               & $\geq 2$               & $\geq 2$               & $\geq 2$               & $\geq 2$               & $\geq 2$               & $\geq 3$               & $\geq 2$       \\
			Lepton charges         & $++/--$                & $++/--$                & $++$                   & $++/--$                & $++/--$                & $++/--$                & $++/--$                & $++/--$                & $++/--$        \\
			$\MET$                 & ${>}0$\GeV             & ${>}30$\GeV            & ${>}30$\GeV            & ${>}120$\GeV           & ${>}50$\GeV            & ${>}50$\GeV            & ${>}120$\GeV           & ${>}50$\GeV            & ${>}0$\GeV     \\
			$\HT$                  & ${>}80$\GeV            & ${>}80$\GeV            & ${>}80$\GeV            & ${>}200$\GeV           & ${>}200$\GeV           & ${>}320$\GeV           & ${>}320$\GeV           & ${>}200$\GeV           & ${>}320$\GeV   \\
			\hline
			Fake BG                & $25 \pm 13$ & $19 \pm 10$ & $9.6 \pm 5.0$ & $0.99 \pm 0.69$ & $4.5 \pm 2.9$ & $2.9 \pm 1.7$ & $0.7 \pm 0.5$ & $0.71 \pm 0.47$ & $4.4 \pm 2.6$  \\
			Charge-flip BG         & $3.4 \pm 0.7$ & $2.7 \pm 0.5$ & $1.4 \pm 0.3$ & $0.04 \pm 0.01$ & $0.21 \pm 0.05$ & $0.14 \pm 0.03$ & $0.04 \pm 0.01$ & $0.03 \pm 0.01$ & $0.21 \pm 0.05$\\
			Rare SM BG             & $11.8 \pm 5.9$ & $10.5 \pm 5.3$ & $6.7 \pm 3.4$ & $1.2 \pm 0.7$ & $3.4 \pm 1.8$ & $2.7 \pm 1.5$ & $1.0 \pm 0.6$ & $0.44 \pm 0.39$ & $3.5 \pm 1.9$  \\
			\hline
			Total BG               & $40 \pm 14$ & $32 \pm 11$ & $17.7 \pm 6.1$ & $2.2 \pm 1.0$ & $8.1 \pm 3.4$ & $5.7 \pm 2.4$ & $1.7 \pm 0.7$ & $1.2 \pm 0.6$ & $8.1 \pm 3.3$  \\
			Event yield            & 43                & 38                & 14                & 1                & 10                & 7                & 1                & 1                & 9              \\
	        \hline
			$N_{{UL}}$ (13\% unc.) & 27.2	&26.0	&9.9	&3.6	&10.8	&8.6	&3.6	&3.7	&9.6 \\
			$N_{{UL}}$ (20\% unc.) & 28.2	&27.2	&10.2	&3.6	&11.2	&8.9	&3.7	&3.8	&9.9 \\
			$N_{{UL}}$ (30\% unc.) & 30.4	&29.6	&10.7	&3.8	&12.0	&9.6	&3.9	&4.0	&10.5 \\
			\hline
			\end{tabular}
		\end{scriptsize}
	\end{table}

\begin{figure}[t]
\begin{center}
\includegraphics[width=0.7\linewidth]{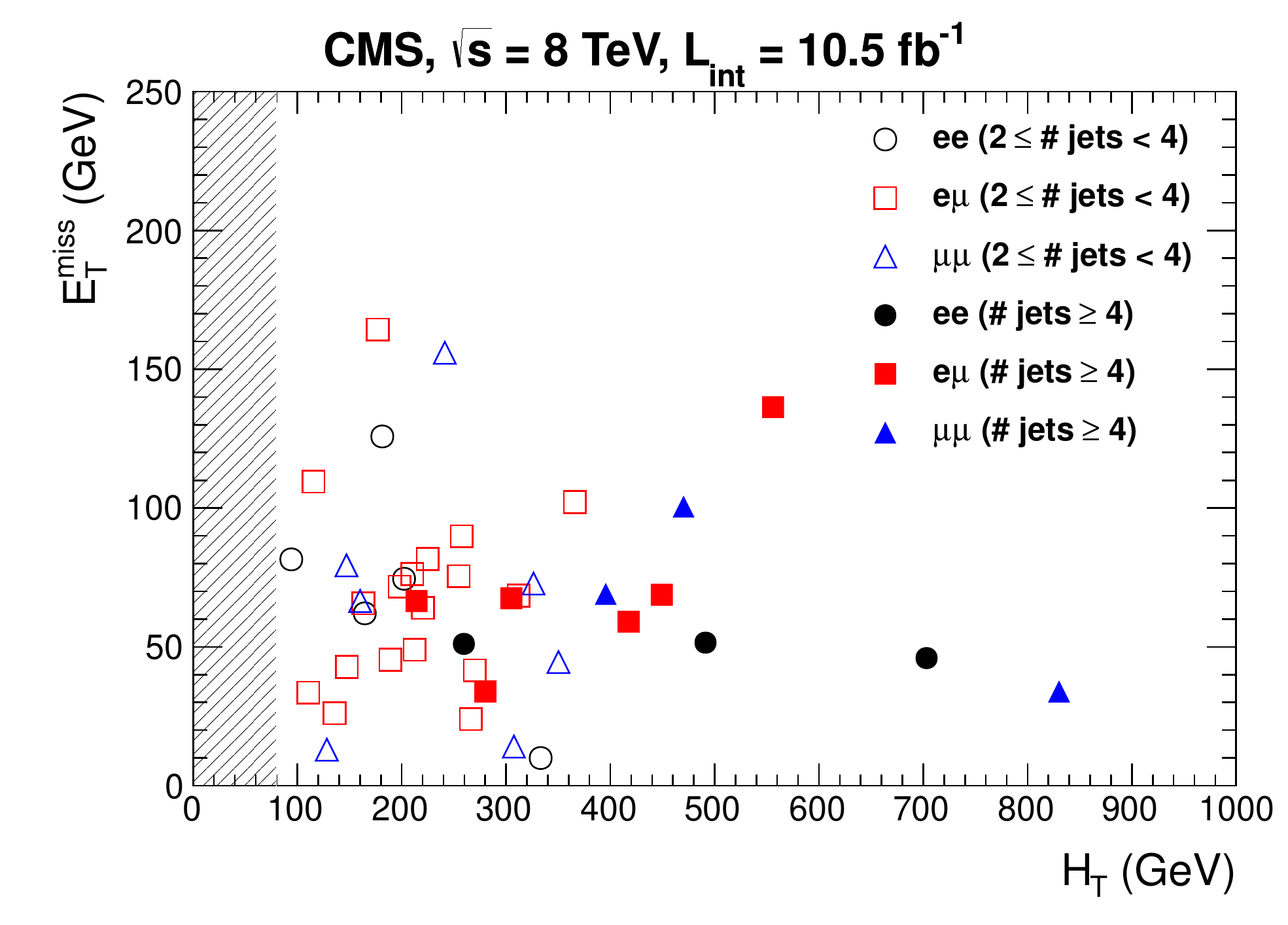}
\includegraphics[width=0.49\linewidth]{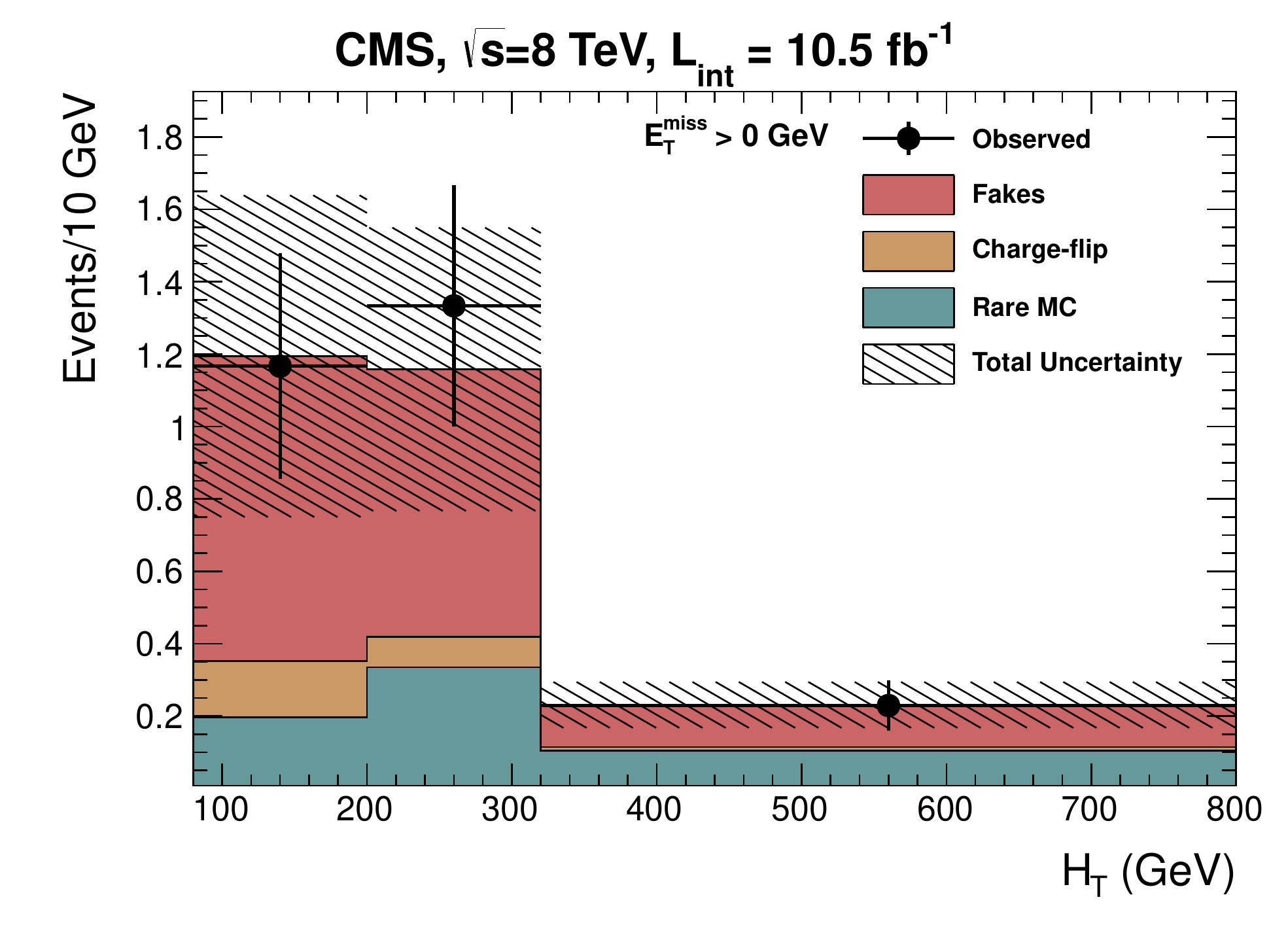}
\includegraphics[width=0.49\linewidth]{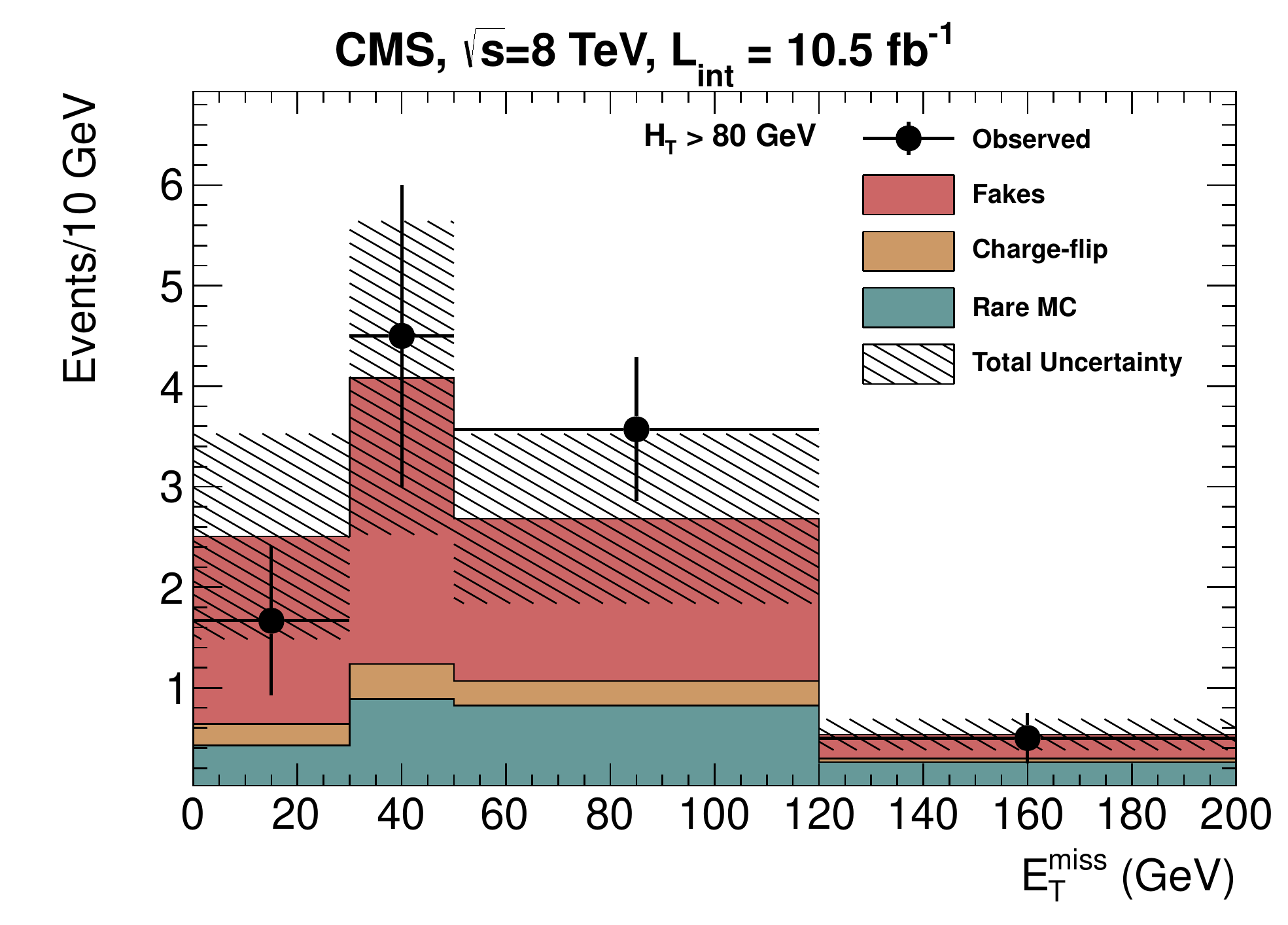}
\caption{\label{fig:htmet}
Top plot: distribution of \MET versus $\HT$ for the 43 events in the
most inclusive signal region (SR0).
Note that the $\geq 2$ jets requirement in SR0 implies $\HT > 80$\GeV.
Bottom-left plot: projection of the top plot on the $\HT$ axis.
Bottom-right plot: projection of the top plot on the \MET axis.
For the one-dimensional distributions, the number of events in each bin is scaled appropriately to reflect units of events per
10\GeV and is compared with the background
predictions, with their uncertainties. The binning of the one-dimensional
distributions is chosen to match the \MET and $\HT$ boundaries of
the different signal regions.}
\end{center}
\end{figure}

The definition of the signal regions and the observations of the search are summarized in
Table~\ref{tab:SRcomb}.  The basic kinematic properties of the
selected
events are shown in
Fig.~\ref{fig:htmet}.
The results are consistent with the background predictions.

Table~\ref{tab:SRcomb} also shows the 95\% confidence level (CL)
observed upper limit ($N_{UL}$)
on the number
of non-SM events, calculated using the CL$_\mathrm{s}$ method~\cite{Junk:1999kv,ATL-PHYS-PUB-2011-011,cls}
under three different assumptions for the
signal efficiency uncertainty.
As can be seen from Table~\ref{tab:SRcomb},
the limits are only weakly
dependent on this uncertainty. The uncertainty on the signal efficiency is discussed
in Section~\ref{sec:eff}.

\section{Efficiencies and associated uncertainties}
\label{sec:eff}

The dilepton triggers require one lepton to have $\pt > 17$\GeV
and the other $\pt > 8$\GeV.
The trigger efficiency, measured using events collected with
hadronic triggers, is
$96 \pm 5\%$ for \Pe \Pe, $93 \pm 5\%$ for \Pe \Pgm, and
$88 \pm 4\%$ for \Pgm \Pgm.  All acceptances, which are calculated
from MC simulation, include a correction factor for the
trigger efficiency.

\begin{figure}[t]
\begin{center}
\includegraphics[width=0.49\linewidth]{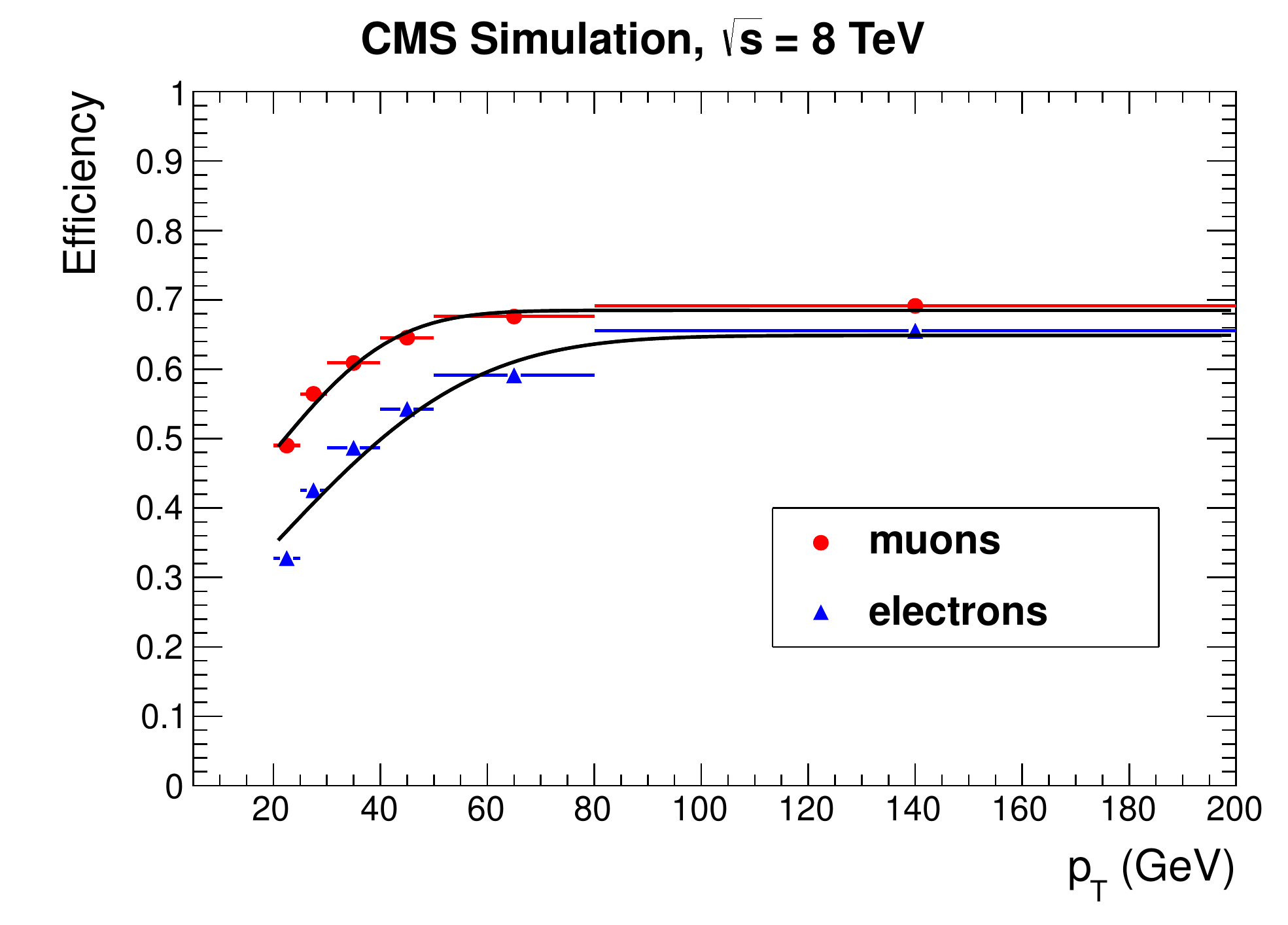}
\includegraphics[width=0.49\linewidth]{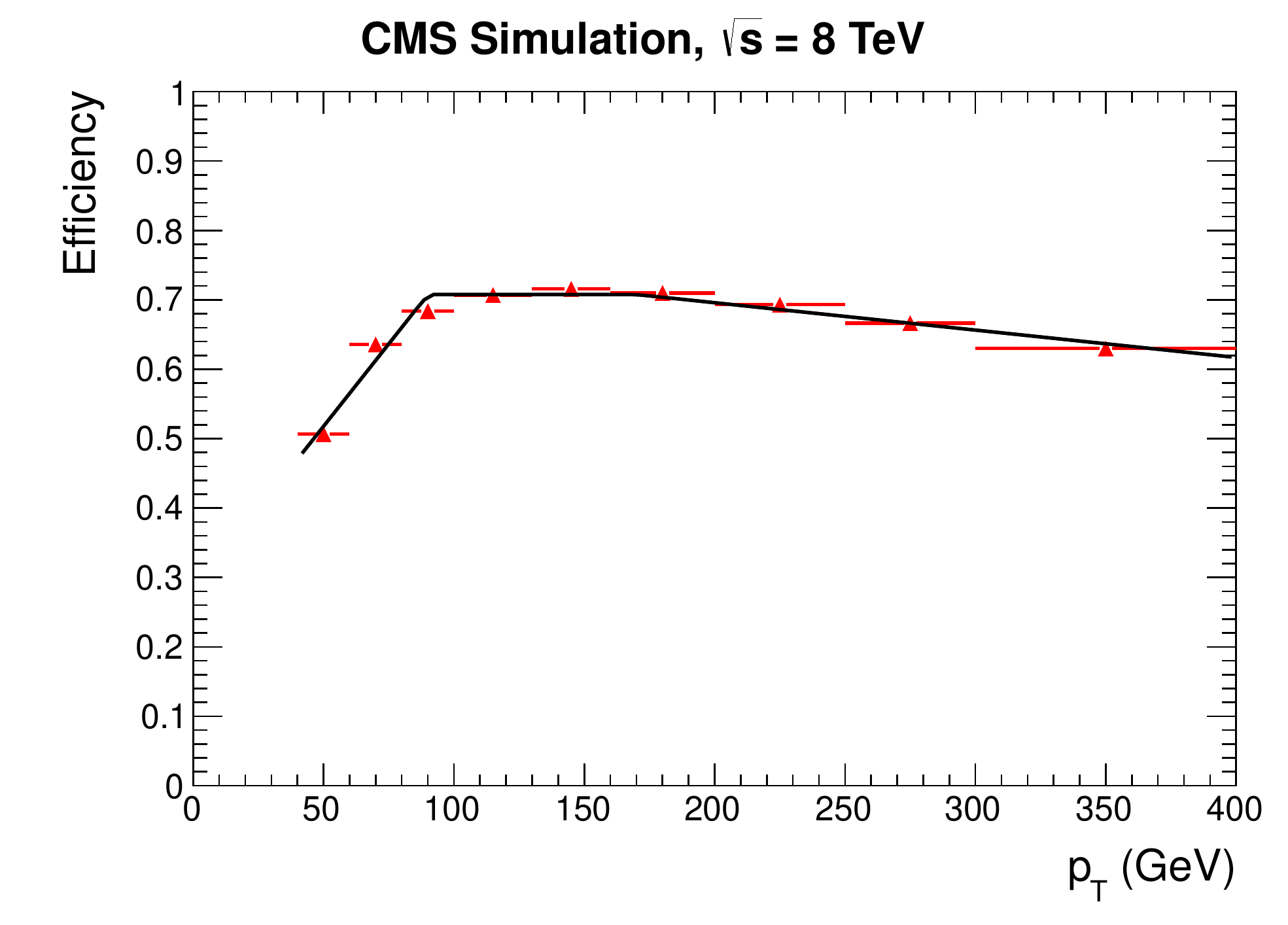}
\caption{\label{fig:efficiencies}
Lepton selection efficiency as a function of $\pt$ (left);
b-tagging efficiency as a function of the b-quark $\pt$ (right).
The superimposed curves represent the efficiency parametrizations discussed in Section~\ref{sec:outreach}.
}
\end{center}
\end{figure}

The lepton selection efficiencies,
determined using a sample of simulated events
from a typical signal scenario involving multiple top quarks in the
final states
arising from gluino decay~\cite{T1tttt,wacker}, are displayed in Fig.~\ref{fig:efficiencies}.
Studies of large samples of $\cPZ \to \Pe \Pe$ and $\cPZ \to \Pgm \Pgm$
events indicate that the
lepton selection efficiencies are reproduced to better than 2\% by the
simulation.   However, leptons in SUSY events have significantly
lower isolation efficiency due to the large amount of hadronic
activity that is expected in these scenarios.  As a result, we
conservatively
assign
an uncertainty of 5\% per lepton to the lepton selection efficiency.

Figure~\ref{fig:efficiencies} also shows the b-tagging efficiency,
obtained
from simulation,
for b quarks with $|\eta| < 2.4$.
The uncertainty on this efficiency is
a function of b-jet $\pt$; it is
4.4\% at $\pt =40\GeV$, 2.7\% at $\pt = 100$\GeV, and
10\% at $\pt = 500$\GeV~\cite{CMS-PAS-BTV-11-003}.

The uncertainty on the energy scale of jets in this analysis
is also a function of $\pt$~\cite{JES}.  It varies between 5\%
and 2\% in the \pt range 40--100\GeV for jets with $|\eta| < 2.4$.

The importance of these effects depends on the
signal region and the model of new physics.
In general, models
with high hadronic activity and large \MET~~are less affected
by the uncertainty on the jet energy scale.

The total uncertainty on the acceptance for the models
described in Section~\ref{sec:models}
is in the 13--20\% range.
Finally, there is a 4.4\% uncertainty on the yield of events from
any new physics model due to the uncertainty in the luminosity
normalization~\cite{CMS-PAS-SMP-12-008}.

\section{Information for model testing}
\label{sec:outreach}

Our
results can be used to confront models of new physics
in an approximate way through generator-level studies
that compare the expected numbers of events with the
upper limits from Table~\ref{tab:SRcomb}.
The prescription to be used is given
in Ref.~\cite{Chatrchyan:2012sa}, Section 7.
The \MET~and \HT turn-on curves
in this analysis are the same as those of Ref.~\cite{Chatrchyan:2012sa}.
However the lepton and b-tagging efficiencies
are slightly different because
of the changes in the underlying selections.  Their current
parametrizations are given below.

The parametrization of the lepton efficiency, shown in
Fig.~\ref{fig:efficiencies}  (left), is

\begin{equation}
\epsilon = \epsilon_{\infty} \erf\left ( \frac{\pt - 20{\GeV}}{\sigma} \right)
        + \epsilon_{20} \left[ 1 - \erf\left ( \frac{\pt -
              20\GeV}{\sigma} \right)\right],
\label{eq:lepeffFitF}
\end{equation}

\noindent with $\epsilon_{\infty} = 0.65$ (0.69),
$\epsilon_{20} = 0.35$ (0.48), and $\sigma = 42$\GeV (25\GeV) for electrons
(muons).
The parametrization of the simulated b-tagging efficiency,
shown in Fig.~\ref{fig:efficiencies} (right),
is $\epsilon = 0.71$ for $90  < \pt < 170$\GeV; at higher
(lower) $\pt$ it decreases linearly with a slope of $-0.0004$
$(-0.0047)\GeV^{-1}$.

\section{Limits on models of new physics}
\label{sec:models}
The results of the search are used
to constrain specific models of new physics.
For each model
considered, limits are derived from the signal region
expected to give the most
stringent limit on the cross section at a given point
in the parameter space of the model.
The event
selection efficiency for a given model is
obtained from MC simulation, and the limits are
calculated including systematic uncertainties on lepton efficiency,
trigger efficiency, luminosity, jet energy scale, and
b-tagging efficiency.  The latter two uncertainties are
evaluated at each point in the parameter space.

The results from SR1 and SR2 are used to set limits on
the cross section for same-sign top-quark pair production,
$\sigma(\Pp \Pp \to \cPqt \cPqt + \cPaqt \cPaqt)$ from SR1, and
$\sigma(\Pp \Pp \to \cPqt \cPqt)$ from SR2.
Here
$\sigma(\Pp \Pp \to \cPqt \cPqt + \cPaqt \cPaqt)$ is shorthand
for the sum
$\sigma(\Pp \Pp \to \cPqt \cPqt)+
\sigma(\Pp \Pp \to \cPaqt \cPaqt)$.
Note that
in most new
physics scenarios
$\Pp \Pp \to \cPaqt \cPaqt$ is suppressed with respect
to $\Pp \Pp \to \cPqt \cPqt$
because of the parton
distribution functions of the proton. These limits are calculated
using simulated $\Pp \Pp \to \cPqt \cPaqt$ events to model the acceptance.
This acceptance,
including branching fractions, is $0.29 \pm 0.04\%$.
We find upper limits
$\sigma(\Pp \Pp \to \cPqt \cPqt + \cPaqt \cPaqt) < 0.87$ pb
and
$\sigma(\Pp \Pp \to \cPqt \cPqt) < 0.30\unit{pb}$ at the 95\% CL;
the median expected limits are 0.72 and 0.37\unit{pb}, respectively.

\begin{figure}[t]
\begin{center}
\includegraphics[width=0.45\linewidth]{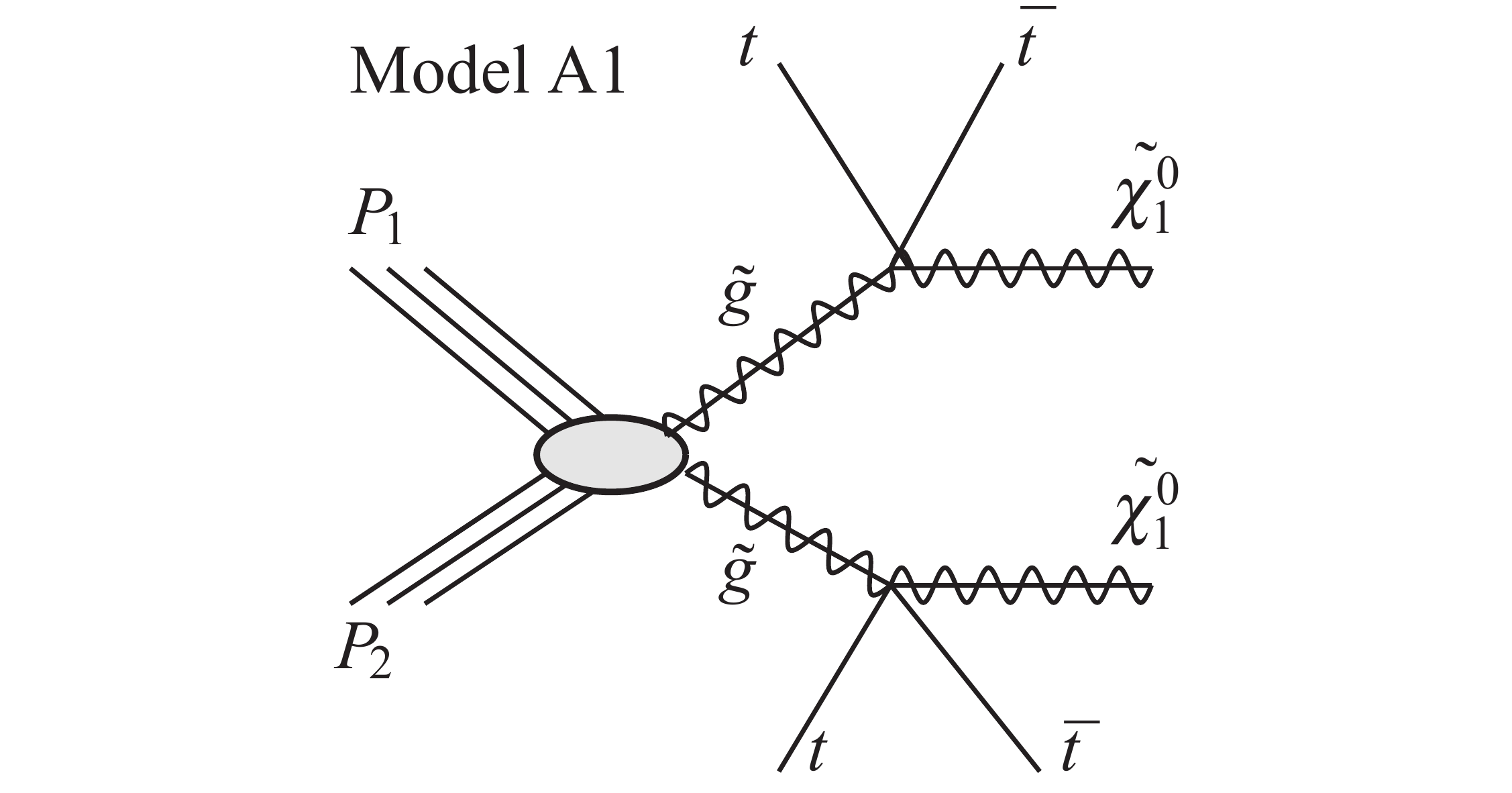}
\includegraphics[width=0.45\linewidth]{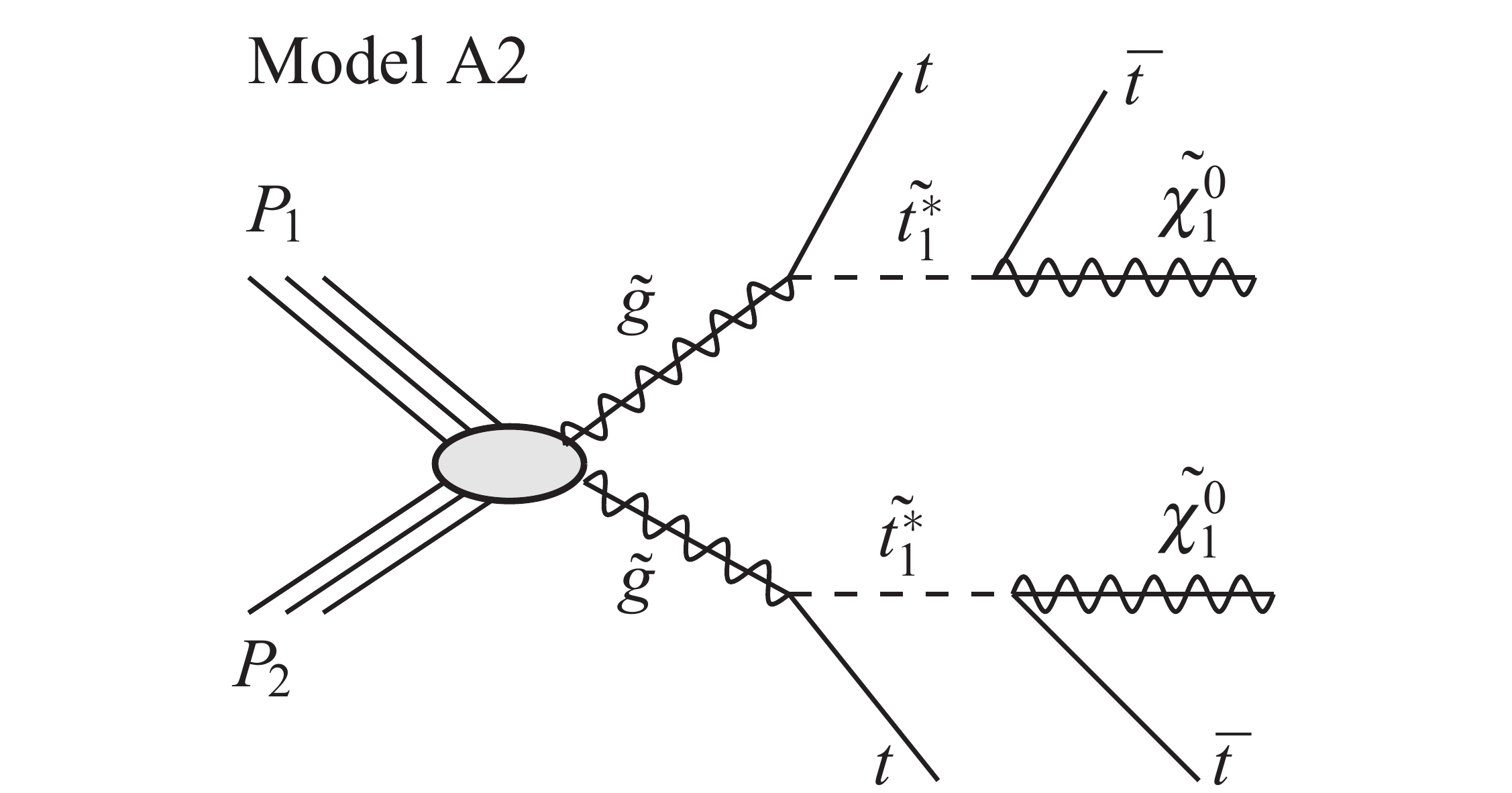}
\includegraphics[width=0.45\linewidth]{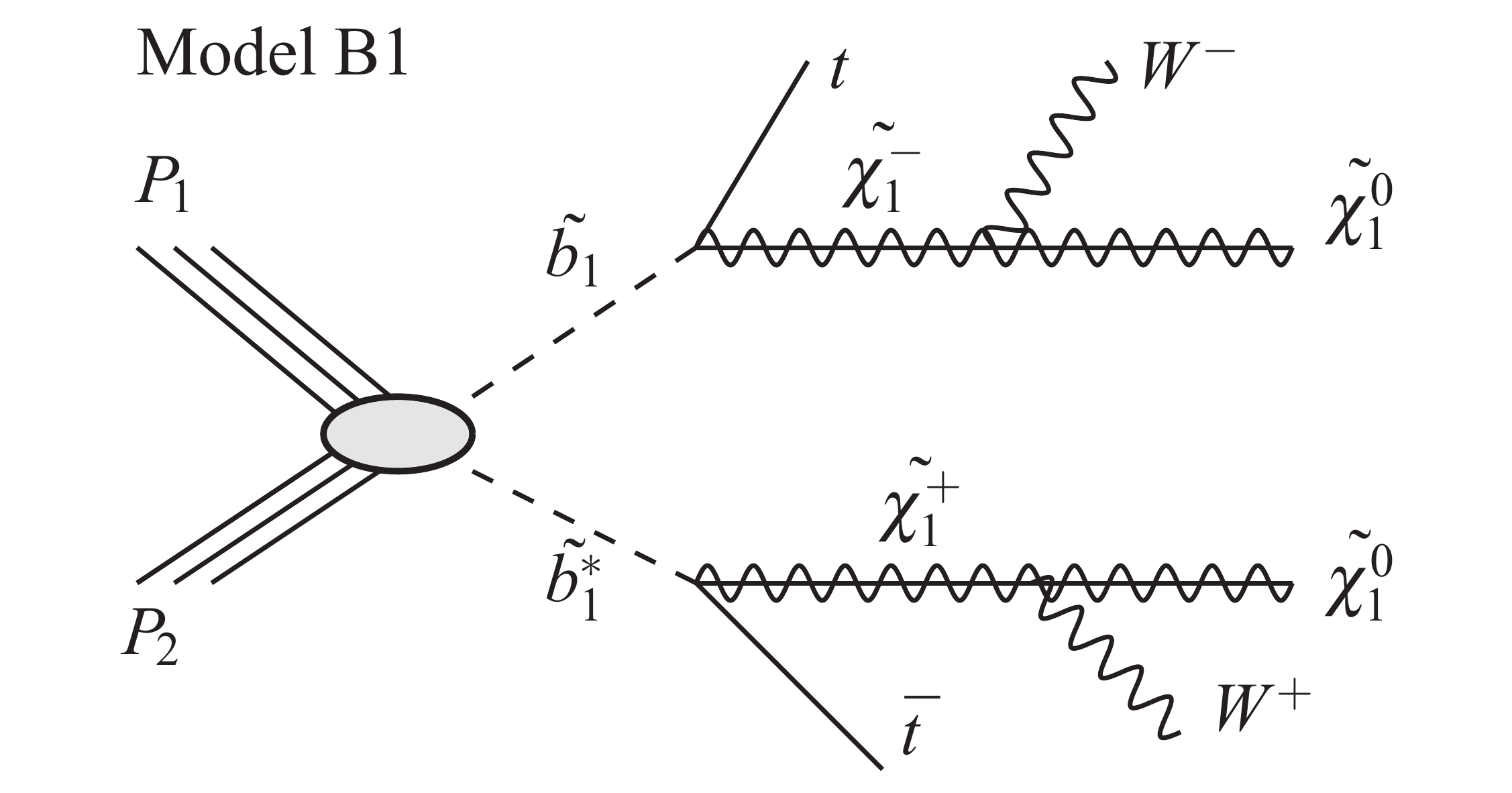}
\includegraphics[width=0.45\linewidth]{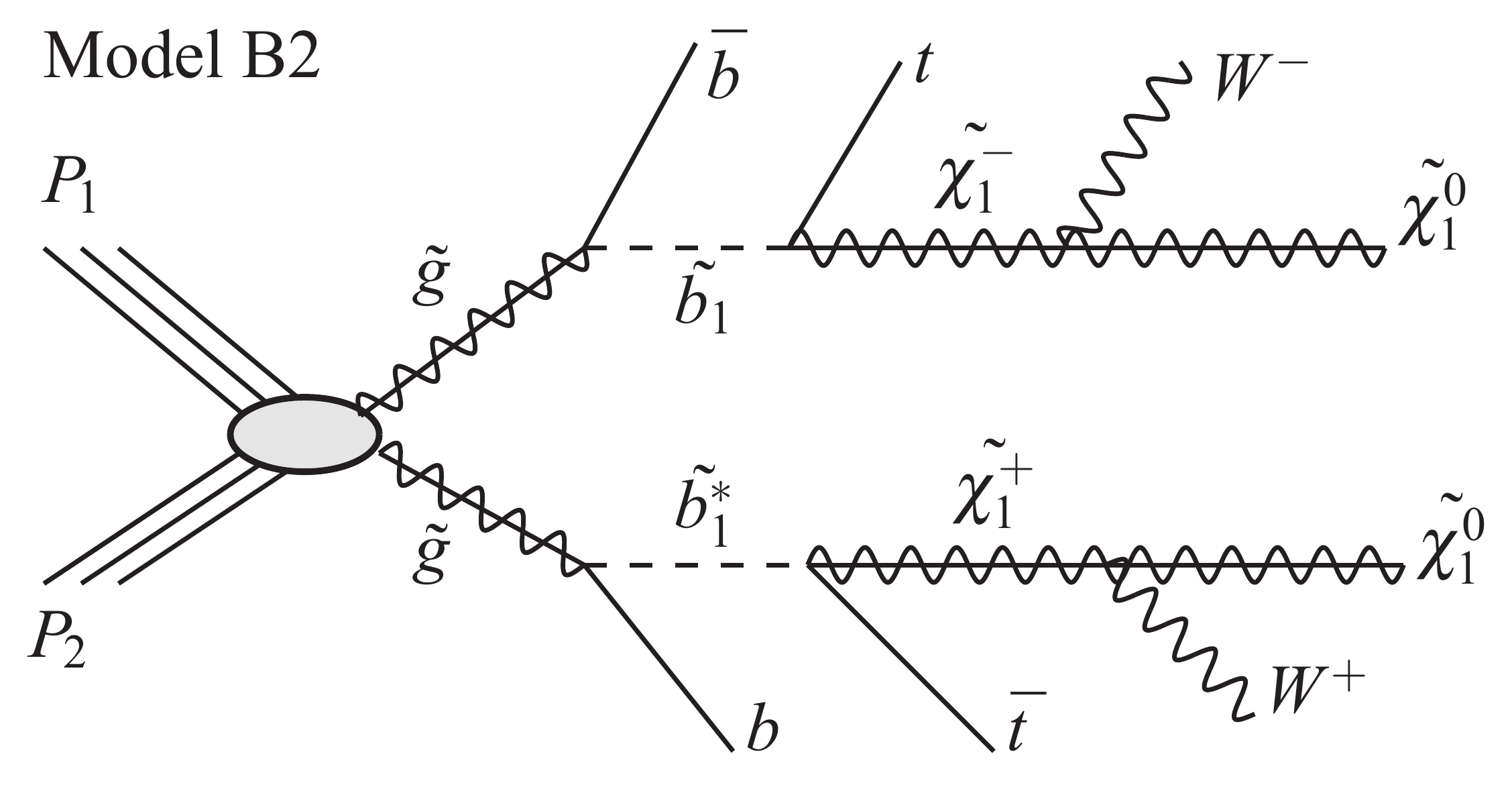}
\caption{Diagrams for the four SUSY models considered (A1, A2, B1, and B2).}
\label{fig:FD}
\end{center}
\end{figure}

Next, we present limits on the parameter spaces of four
R-parity-conserving SUSY models
with third-generation squarks.
The decay chains under consideration are shown schematically in
Fig.~\ref{fig:FD}.

Scenarios A1 and A2 represent models of gluino pair production
resulting in the $\cPqt \cPqt \cPaqt \cPaqt \chiz_1 \chiz_1$ final
state, where $\chiz_1$ is the lightest neutralino~\cite{stopVirtual,stopVirtualPRD,T1tttt,wacker,naturalness4}.
In model A1, the gluino undergoes a three-body decay
$\sGlu \to \ttbar \chiz_1$ mediated by an off-shell top squark.
In model
A2, the gluino decays to a top quark and and an anti-top squark,
with the on-shell anti-squark
further decaying into an anti-top quark and a neutralino.

Models B1 and B2 have bottom squarks decaying
as $\sBot_1 \to \cPqt \chim_1$ and $\chim_1 \to \PWm \chiz_1$.
Model B1 is a model of bottom-squark pair production, followed
by one of the most likely decay modes of the bottom squark; model B2
consists of gluino pair production followed by $\sGlu \to \sBot_1 \cPaqb$.
The gluino decay modes in models A1 and A2 would be dominant
if the top squark is the lightest supersymmetric quark.
Conversely, if the bottom squark is lightest, the decay mode
in model B2 would be the most probable.

\begin{figure}[tp]
\begin{center}
\includegraphics[width=0.49\linewidth]{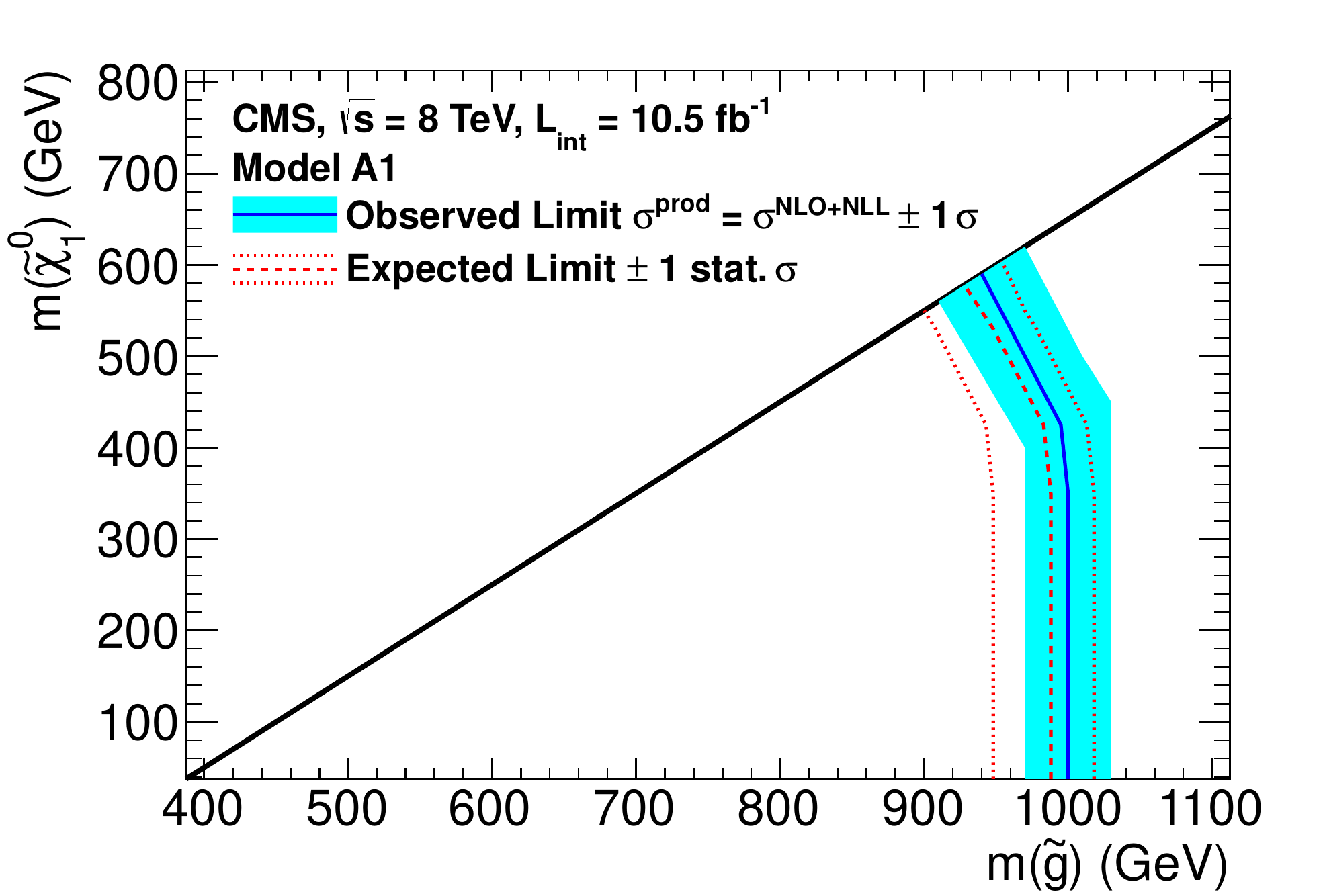}
\includegraphics[width=0.49\linewidth]{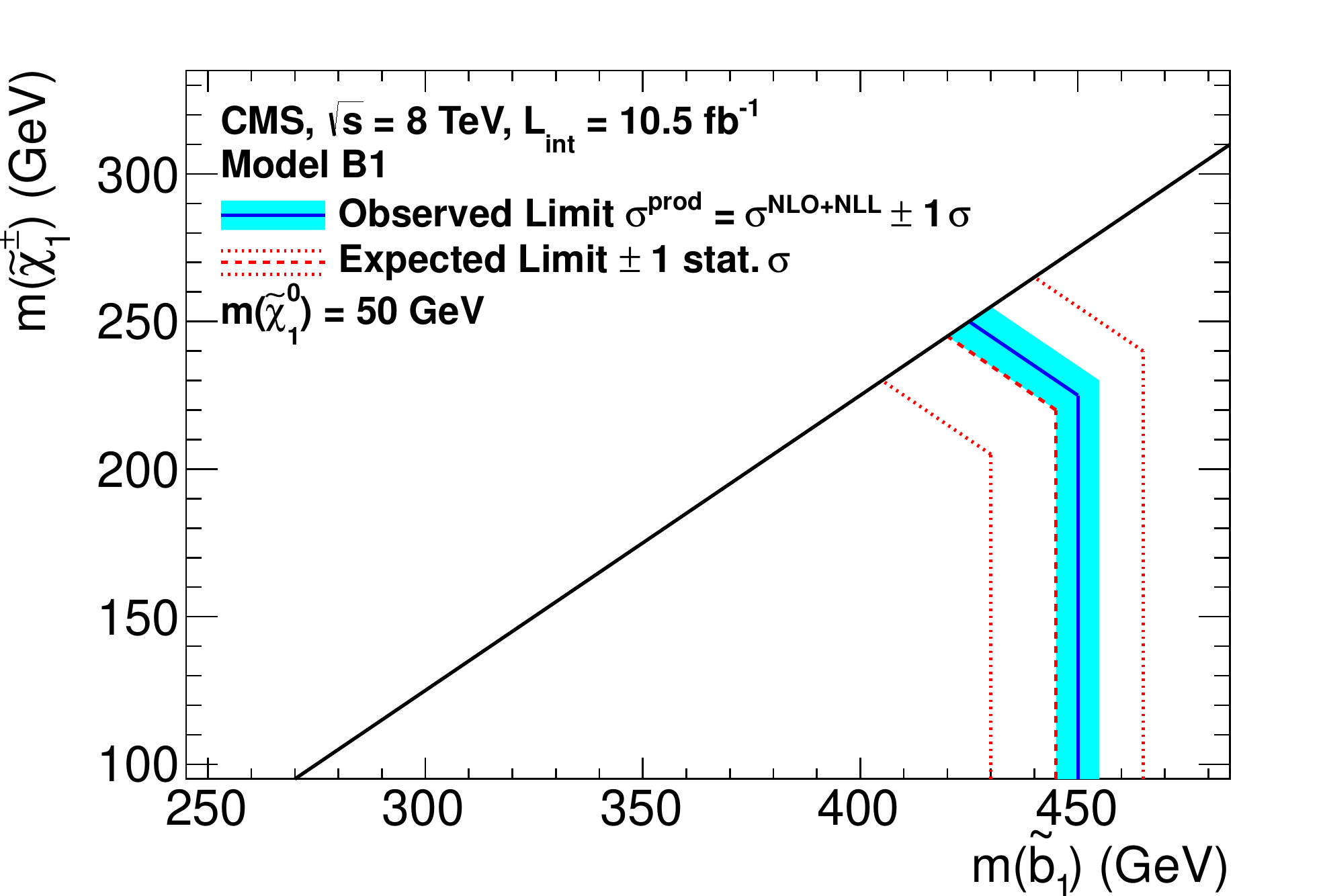}
\includegraphics[width=0.49\linewidth]{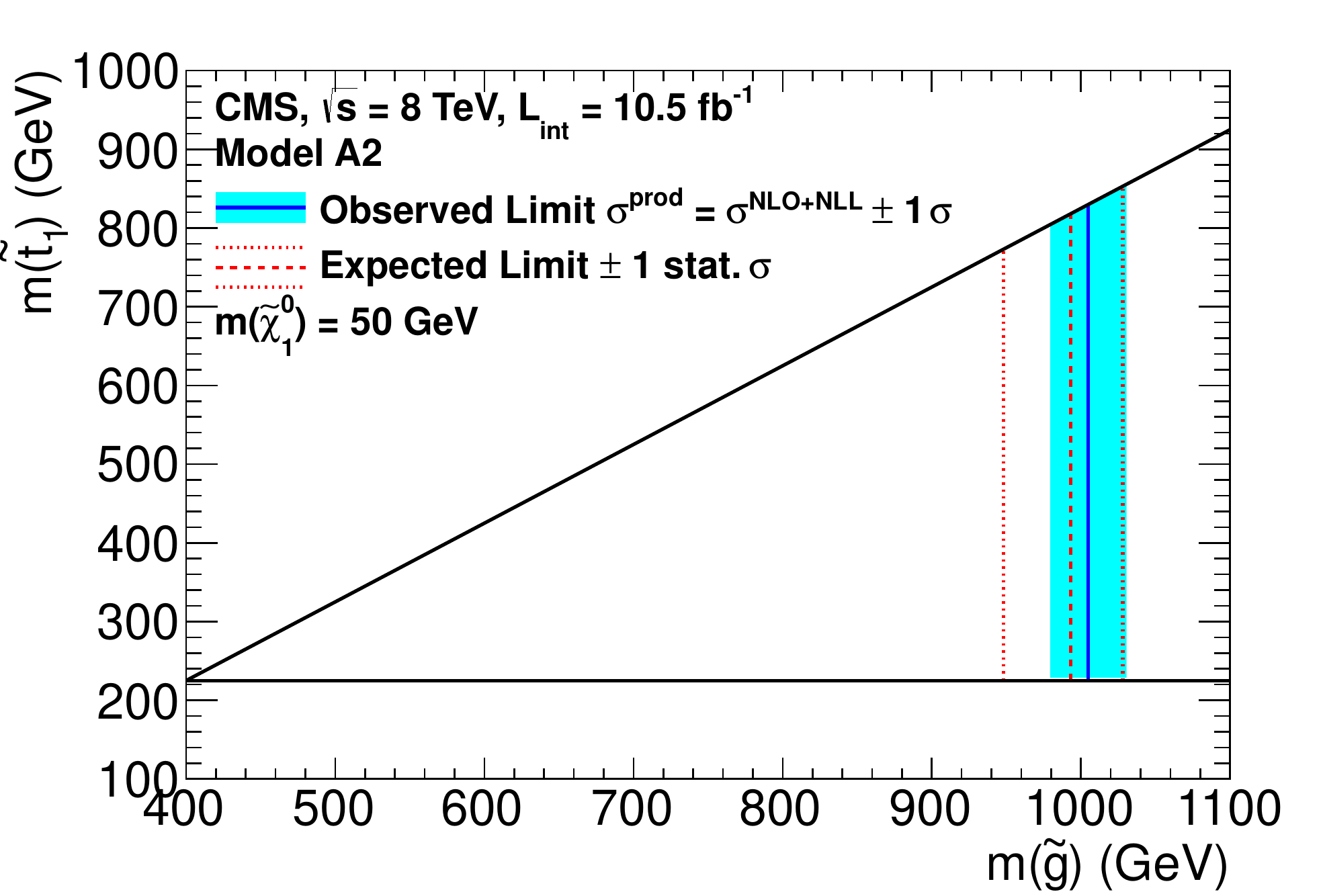}
\includegraphics[width=0.49\linewidth]{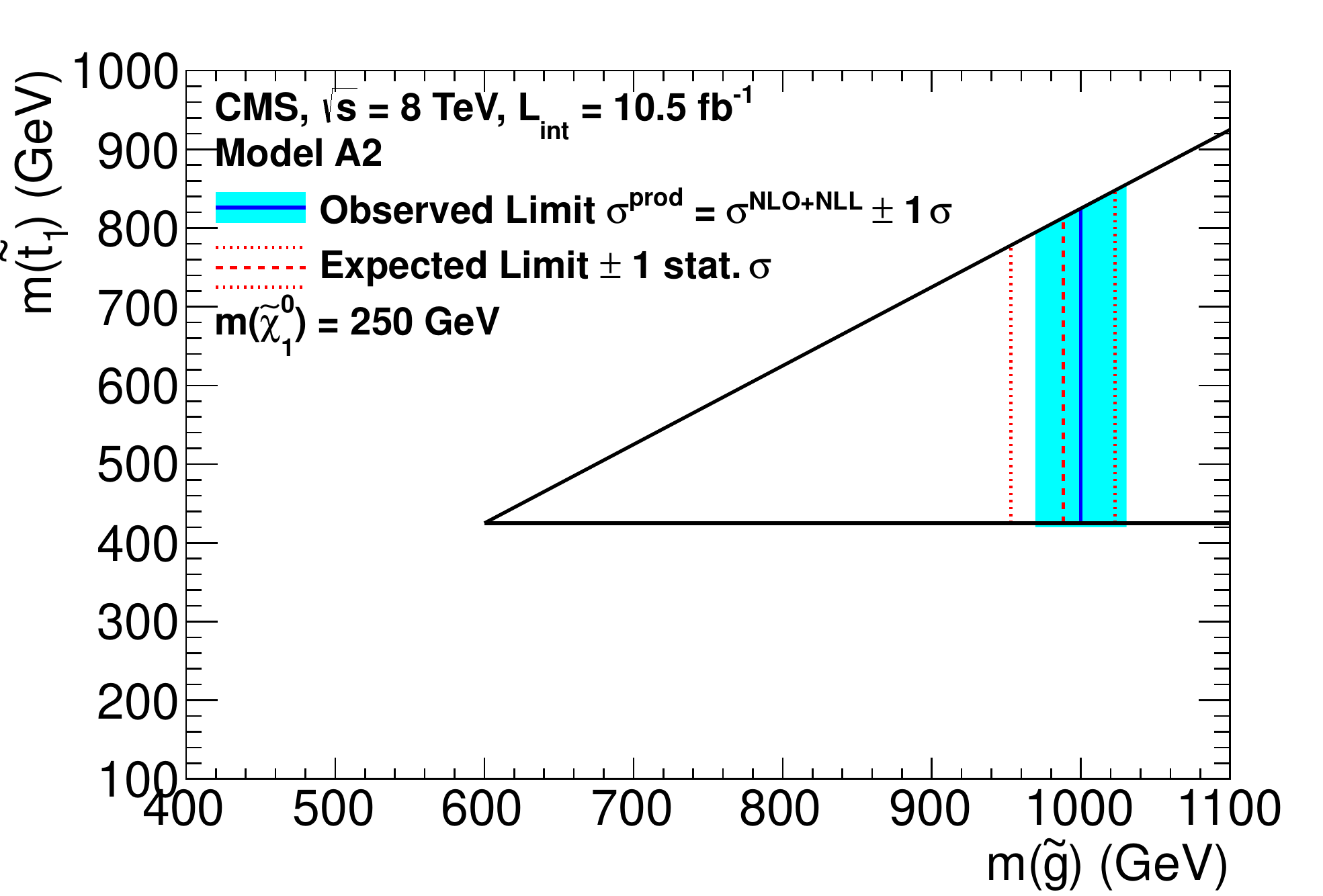}
\includegraphics[width=0.49\linewidth]{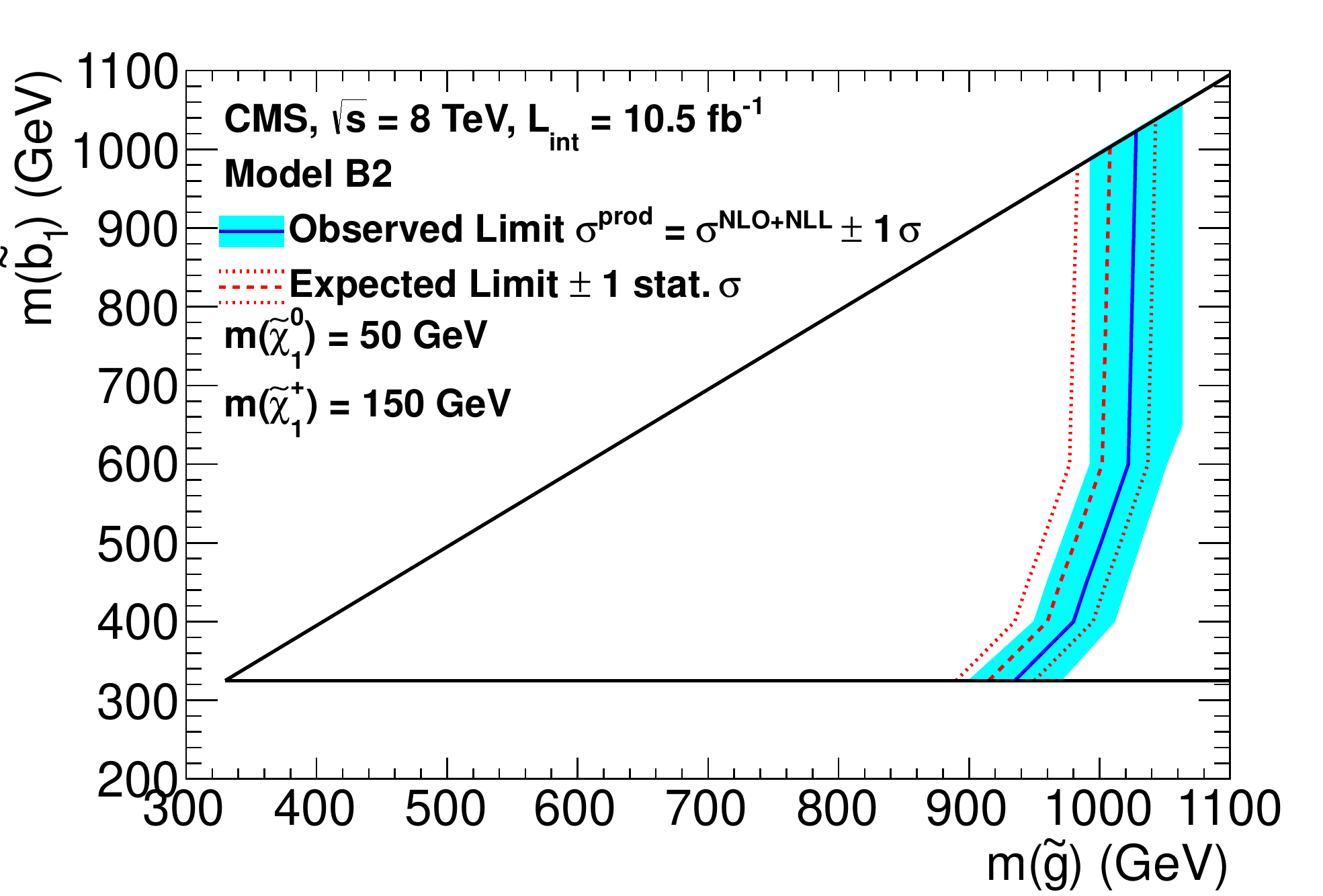}
\includegraphics[width=0.49\linewidth]{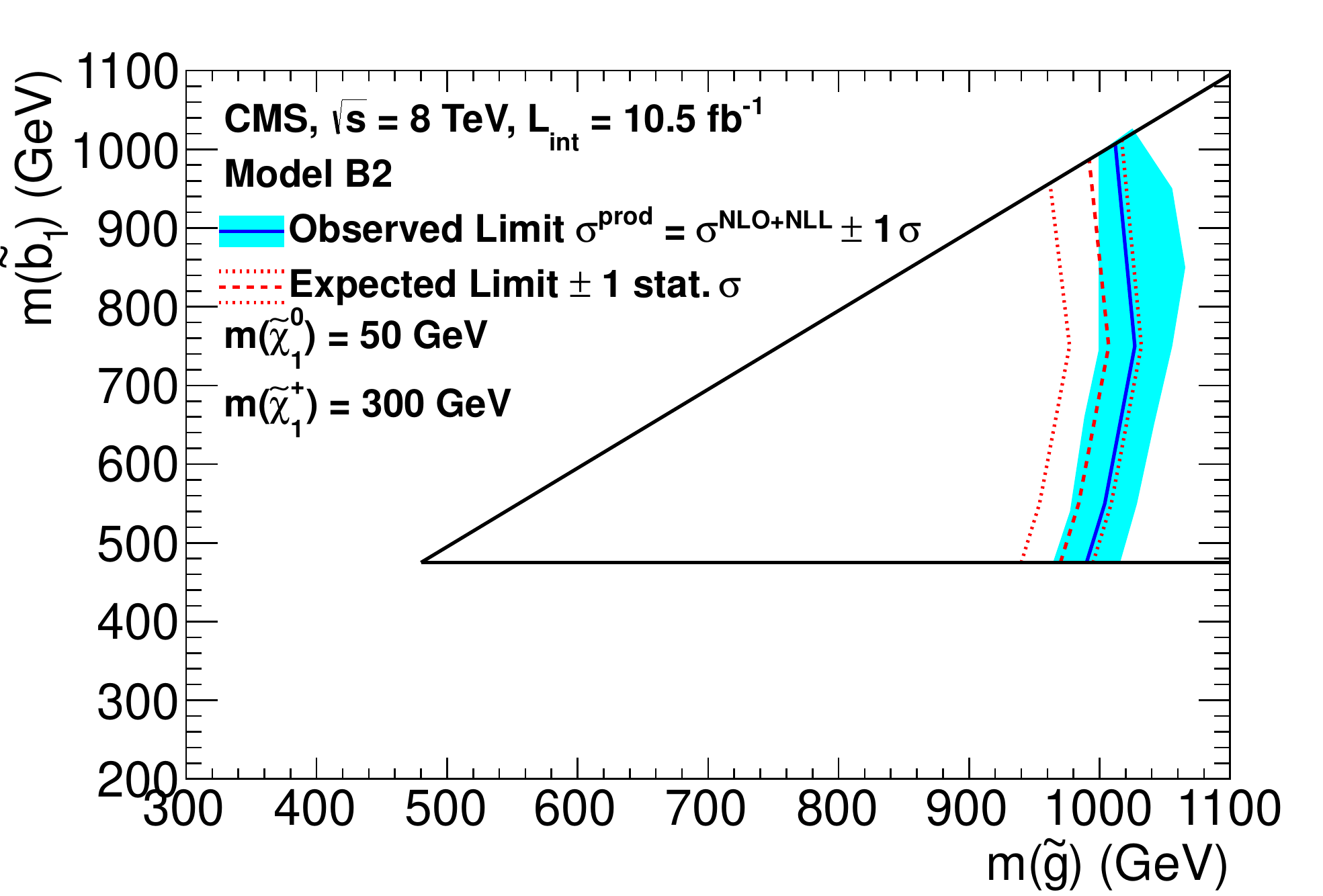}
\caption{Exclusion regions at 95\% CL in the planes of
$m(\chiz_1)$ vs.\ $m(\sGlu)$ (model A1),
$m(\chim_1)$ vs.\ $m(\sBot_1)$ (model B1),
$m(\sTop_1)$ vs.\ $m(\sGlu)$ (model A2), and
$m(\sBot_1)$ vs.\ $m(\sGlu)$ (model B2).
Models A2, B1, and B2 have more than two mass parameters, and cannot
be fully represented in a two dimensional plot.  The assumed values
of the additional mass parameters are indicated in the plots.
The black lines represent the kinematic boundaries of the models.
The excluded regions are those within the kinematic boundaries and to
the left of the bands.
The effects of the theoretical uncertainties on the
next-to-leading-order
plus next-to-leading-log calculations of the production
cross sections~\cite{Kramer:2012bx}
are indicated by the shaded bands; the expected limits and their $\pm$1
standard-deviation variations are also shown.
\label{fig:lipstick}}
\end{center}
\end{figure}

Excluded regions in the parameter space of the four models are shown in Fig.~\ref{fig:lipstick}.
For the gluino-initiated models (A1, A2, and B2), we exclude gluinos
with masses below around 1\TeV, with only a small dependence on the
details of the models.  This is because the limits are driven by the
common gluino pair production cross section.
In the case of bottom squark pair production (model
B1) we set a limit $m(\sBot_1) > 450$\GeV at 95\% CL.

These models are also probed by other CMS new physics searches
in different decay modes, although other searches have so far
been interpreted in the context of model A1 but not A2, B1, or B2.
For model A1 the limits given here are
complementary
to the limits from the single-lepton and  ``razor''
searches~\cite{maria}: less stringent at low
$m(\chiz_1)$ but more stringent at high $m(\chiz_1)$.
A similar conclusion applies to model A2, since the
final state is the same.
Comparable limits for model A1, as well as for similar models with
top and bottom quarks from gluino decays, have been
reported by the ATLAS
collaboration~\cite{ATLAS:2012ai,ATLAS:2012ah,ATLAS:2012pq}.
In the case of bottom-squark pair production, the ATLAS collaboration
has reported a limit $m(\sBot_1) > 390$\GeV, but assuming the decay
mode $\sBot_1 \to \cPqb \chiz_1$ instead of the model B1 mode
$\sBot_1 \to \cPqt \chim_1$~\cite{Aad:2011cw}.

\section{Summary}
\label{sec:conclusion}

In summary, we have presented results of a search for
same-sign dileptons with bottom-quark jets using the
CMS detector at the LHC, based on
a 10.5\fbinv data sample of pp collisions at $\sqrt{s}$ = 8\TeV.
No significant deviations from the standard model expectations are
observed.

The results are used to set upper limits
on the same-sign top-pair production cross section
$\sigma(\Pp \Pp \to \cPqt \cPqt + \cPaqt \cPaqt) < 0.87$ pb
and
$\sigma(\Pp \Pp \to \cPqt \cPqt) < 0.30\unit{pb}$ at the 95\%
confidence level.

We also exclude
gluinos with masses up to approximately 1\TeV if they
decay exclusively into top or bottom squarks,
and we place a
lower limit on the bottom-squark mass of 450\GeV.
Our results assume
that the top
and bottom squarks decay as
$\sTop_1 \to \cPqt \chiz_1$
and
$\sBot_1 \to \cPqt \chim_1$, respectively.
In the latter case we have also assumed
$\chim_1 \to \PWm \chiz_1$.

\section*{Acknowledgements}
We wish to congratulate our colleagues in the CERN accelerator departments for the excellent performance of the LHC machine.
We thank the technical and administrative staff at CERN and other CMS institutes, and acknowledge support from:
FMSR (Austria); FNRS and FWO (Belgium); CNPq, CAPES, FAPERJ, and FAPESP (Brazil);
MES (Bulgaria); CERN; CAS, MoST, and NSFC (China); COLCIENCIAS (Colombia);
MSES (Croatia); RPF (Cyprus); Academy of Sciences and NICPB (Estonia); Academy of Finland, MEC, and HIP (Finland);
CEA and CNRS/IN2P3 (France); BMBF, DFG, and HGF (Germany); GSRT (Greece); OTKA and NKTH (Hungary); DAE and DST (India);
IPM (Iran); SFI (Ireland); INFN (Italy); NRF and WCU (Korea); LAS (Lithuania);
CINVESTAV, CONACYT, SEP, and UASLP-FAI (Mexico); MSI (New Zealand); PAEC (Pakistan);
MSHE and NSC (Poland); FCT (Portugal); JINR (Armenia, Belarus, Georgia, Ukraine, Uzbekistan);
MON, RosAtom, RAS and RFBR (Russia); MSTD (Serbia); MICINN and CPAN (Spain);
Swiss Funding Agencies (Switzerland); NSC (Taipei); TUBITAK and TAEK (Turkey); STFC (United Kingdom); DOE and NSF (USA).

\bibliography{auto_generated}   

\cleardoublepage \appendix\section{The CMS Collaboration \label{app:collab}}\begin{sloppypar}\hyphenpenalty=5000\widowpenalty=500\clubpenalty=5000\textbf{Yerevan Physics Institute,  Yerevan,  Armenia}\\*[0pt]
S.~Chatrchyan, V.~Khachatryan, A.M.~Sirunyan, A.~Tumasyan
\vskip\cmsinstskip
\textbf{Institut f\"{u}r Hochenergiephysik der OeAW,  Wien,  Austria}\\*[0pt]
W.~Adam, E.~Aguilo, T.~Bergauer, M.~Dragicevic, J.~Er\"{o}, C.~Fabjan\cmsAuthorMark{1}, M.~Friedl, R.~Fr\"{u}hwirth\cmsAuthorMark{1}, V.M.~Ghete, N.~H\"{o}rmann, J.~Hrubec, M.~Jeitler\cmsAuthorMark{1}, W.~Kiesenhofer, V.~Kn\"{u}nz, M.~Krammer\cmsAuthorMark{1}, I.~Kr\"{a}tschmer, D.~Liko, I.~Mikulec, M.~Pernicka$^{\textrm{\dag}}$, D.~Rabady\cmsAuthorMark{2}, B.~Rahbaran, C.~Rohringer, H.~Rohringer, R.~Sch\"{o}fbeck, J.~Strauss, A.~Taurok, W.~Waltenberger, C.-E.~Wulz\cmsAuthorMark{1}
% ...
\vskip\cmsinstskip
\dag:~Deceased\\
1:~~Also at Vienna University of Technology, Vienna, Austria\\
2:~~Also at CERN, European Organization for Nuclear Research, Geneva, Switzerland\\
% ...

\end{sloppypar}
\end{document}